\documentclass{aa}  

\usepackage{graphicx}
\usepackage{txfonts}
\usepackage{hyperref}
\usepackage{etoolbox}

\makeatother

\begin{document}

   \title{External gas accretion provides a fresh gas supply to the active S0 galaxy NGC 5077}

   \author{S. I. Raimundo
          \inst{1,2,3}
          }

   \institute{$^{1}$ DARK, Niels Bohr Institute, University of Copenhagen, Lyngbyvej 2, 2100 Copenhagen, Denmark\\
   $^{2}$ Department of Physics and Astronomy, University of California, Los Angeles, CA 90095, USA\\
   $^{3}$ Department of Physics \& Astronomy, University of Southampton, Highfield, Southampton SO17 1BJ, UK\\
   \email{sandra.raimundo@nbi.ku.dk}
             }
 
  \abstract
   {In early type galaxies, externally accreted gas is thought to be the main source of gas replenishment at late times. We use MUSE integral field spectroscopy data to study the active S0 galaxy NGC 5077, known to have disturbed dynamics, indicative of a past external interaction. We confirm the presence of a stellar kinematically distinct core with a diameter of 2.8 kpc, counter-rotating with respect to the main stellar body of the galaxy. We find that the counter-rotating core consists of an old stellar population, not significantly different from the rest of the galaxy. 
   The ionised gas is strongly warped and extends out to 6.5 kpc in the polar direction and in a filamentary structure. The gas dynamics is complex, with significant changes in the position angle as a function of radius. The ionised gas line ratios are consistent with LINER excitation by the AGN both in the nucleus and at kiloparsec scales. We discover a nuclear outflow with projected velocity $V \sim 400$ km/s, consistent with a hollow outflow cone intersecting the plan of the sky.
The properties of the misaligned gas match predictions from numerical simulations of misaligned gas infall after a gas-rich merger. The warp and change in the gas orientation as a function of radius are consistent with gas relaxation due to stellar torques, that are stronger at small radii where the gas aligns faster than in the outer regions, driving gas to the nucleus.
The stellar and gas dynamics indicate that NGC 5077 has had at least two external interactions, one that resulted in the formation of the counter-rotating core followed by late time external gas accretion. NGC 5077 illustrates the importance of external interactions in the replenishment of the galaxy gas reservoir and the nuclear gas content available for black hole fuelling.
   }

   \keywords{AGN --
                galaxy --
                NGC 5077
               }

   \maketitle
%

\section{Introduction}
Counter-rotating or strongly misaligned structures of gas or stars are fairly common in early-type galaxies (e.g.  \citealt{franx&illingworth88}, \citealt{kannapan&fabricant01}, \citealt{falcon-barroso04}, \citealt{mcdermid06},  \citealt{silchenko09}, \citealt{krajnovic11}, \citealt{raimundo17}, \citealt{johnston18}). Stellar kinematically distinct cores (KDCs), i.e. misaligned cores of stars that show an abrupt change of more than 30$^{\circ}$ in their kinematic axis \citep{krajnovic11}, are found in $\sim13 \%$ of the ATLAS 3D sample of early type galaxies (\citealt{ebrova20} and references therein). The presence of misaligned ionised or molecular gas (kinematic misalignment angles of more than 30$^{\circ}$ with respect to the stellar main body rotation), is also common. Out of the early-type galaxies with gas detections, which correspond to $\sim 70 \%$ of galaxies in the ATLAS 3D sample, 41$\%$ show misaligned gas \citep{davis11}. S0 galaxies in particular, tend to have a large fraction of gas in counter-rotation ($\sim 20 - 40 \%$, e.g. \citealt{kuijken96}, \citealt{pizzella04}, \citealt{bureau&chung06}, \citealt{katkov14}). Counter-rotation corresponds to a kinematic misalignment of 180$^{\circ}$ between the stellar and gas kinematic angles and is likely caused by the stellar discs, which tend to align externally accreted gas either in co-rotation or counter-rotation (e.g. \citealt{vandevoort15}).
The presence of misaligned structures is a clear indication that the galaxy underwent an external accretion event (\citealt{bertola92}, \citealt{davis&bureau16}), such as a major or minor merger, gas transfer from a neighbour galaxy, gas accretion from the immediate environment triggered by a flyby, or by the accretion of primordial gas through large scale filaments. This is because the gas that originates from stellar mass loss is expected to co-rotate with the stars. To change the angular momentum of a large mass of gas and create large kinematic misalignments or counter-rotation, one would need a significant amount of energy, unlikely to be met by secular processes in the galaxy (e.g. \citealt{dumas07}, \citealt{davies14}). For galaxies with a low content of native gas, such as some early-type galaxies, the external accretion of gas may provide the necessary fuel to promote one or more episodes of star formation (e.g. \citealt{kaviraj14}, \citealt{davis&young19}) or to fuel an Active Galactic Nucleus (AGN) \citep{raimundo17}. It has been shown from numerical simulations that the presence of misaligned or counter-rotating structures are associated with the flow of gas towards the nucleus and the potential fuelling of the AGN (e.g. \citealt{negri14}, \citealt{capelo&dotti17}, \citealt{taylor18}). One of the best observational examples of such phenomenon is the active galaxy MCG--6-30-15 where the entire reservoir of gas available for AGN fuelling consists of counter-rotating gas (\citealt{raimundo17}). The detection of counter-rotating molecular gas in the nucleus of this galaxy indicates that an external accretion event was able to drive gas to the innermost regions of the galaxy, which fulfils two of the requirements thought to be necessary for AGN fuelling: the presence of gas and the dynamical mechanisms necessary to transport the gas to the black hole (e.g. \citealt{storchi-bergmann19} and references therein). 

MCG--6-30-15 is until now the only active galaxy for which a detailed study of the stellar counter-rotating core, counter-rotating gas dynamics and the impact on black hole fuelling has been carried out (\citealt{raimundo13}, \citealt{raimundo17}). Identifying galaxies with counter-rotating gas is mostly done serendipitously (e.g. \citealt{raimundo13}). We have started a study of counter-rotating gas in the nuclei of galaxies to investigate the hypothesis that external gas accretion can replenish the AGN fuelling reservoir at scales of tens to hundreds of parsecs. The target of the present work, the galaxy NGC 5077, is known to have an active nucleus with characteristics of low-ionisation nuclear emission-line regions (LINERs) and to have ionised gas that is kinematically misaligned with respect to the main stellar body of the galaxy (e.g. \citealt{bertola91}). The particular stellar/gas configuration and the presence of the ionising radiation from the AGN provides an opportunity to study the external accretion event and the AGN properties in this galaxy.

In this work we use for the first time integral field spectroscopy data of NGC 5077 to determine the full extent, dynamics and excitation mechanisms of the ionised gas and to constrain the properties of the external accretion event.

\section{Target and data analysis}
\label{sec:data}
NGC 5077 is an early-type galaxy classified as S0 according to RSA (\citealt{sandage&tammann87}, \citealt{temi07}) and the brightest of a small group of 8 galaxies (e.g. \citealt{sanchez-portal04}, \citealt{tal09}) in a group environment with density $\rho = 0.23$ galaxies/Mpc$^{3}$ \citep{annibali10}. It is located at z $=$ 0.00936, corresponding to a physical scale of 220 pc/arcsec and a luminosity distance of 46.4 Mpc. It has a low level AGN with LINER properties (\citealt{zhang08}, \citealt{annibali10}).
The central black hole has a large mass, and is one of the limited number of supermassive black holes with a dynamically measured mass: M$_{\rm BH} = 8.55^{+4.35}_{-4.48} \times 10^{8}$ M$_{\odot}$ (\citealt{defrancesco08}, for the cosmology parameters used in \citealt{kormendy&ho13}).
There are several indications that NGC 5077 underwent an external accretion event: the overall ionised gas distribution has a polar configuration, approximately orthogonal to the stellar major axis \citep{bertola91}. There is also a counter-rotating stellar core in the nucleus\citep{caon00} and faint dust lanes and filaments observed (\citealt{tran01}, \citealt{simoeslopes07}) pointing towards a past external interaction. 

\begin{figure*}
\hspace{-0.6cm}\includegraphics[width=20cm]{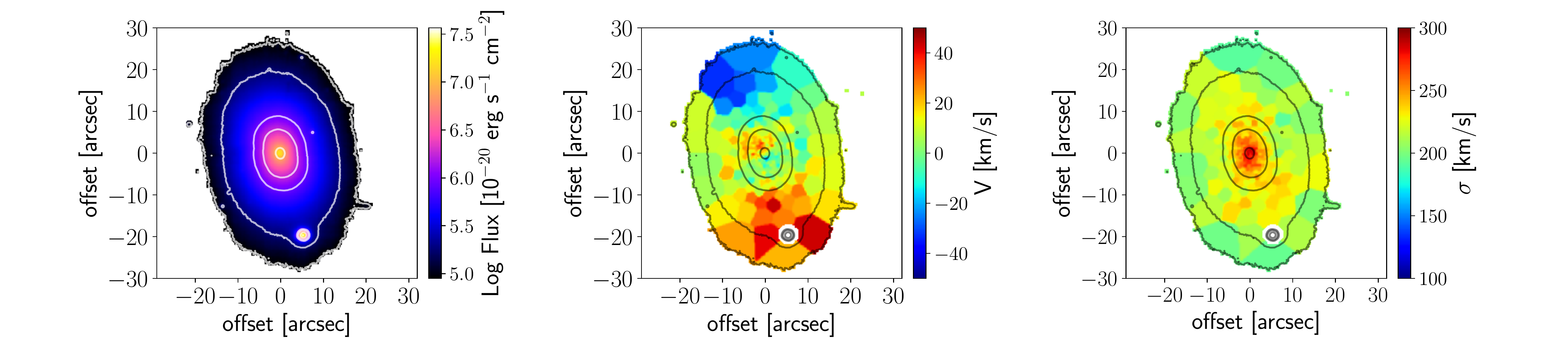}\\
\caption{Two-dimensional maps of integrated light, stellar velocity and stellar velocity dispersion. Left panel: map of integrated light within the MUSE spectral range. The isophotes are shown as contours of 50\%, 10\%, 5\%, 1\% and 0.5\% of the peak flux value. Regions with lower flux were masked out of the image. Middle panel: line of sight stellar velocity. Right panel: stellar velocity dispersion. North is up, east is left and the origin is defined as the position of the AGN. 10 arcsec corresponds to 2.2 kpc. The point source south of the nucleus is a foreground star that was masked out in the centre and right panels.}
\label{stellar_maps}
\end{figure*}

We use archival data on NGC 5077 from the MUSE optical integral field spectrograph on the Very Large Telescope (VLT) (Programme ID: 094.B-0298) to investigate the stellar and gaseous properties of the galaxy. The data were processed using the command line tool \textsc{esorex} (version 3.12.3) and the ESO recipes for the MUSE pipeline (version 2.2.0) \citep{weilbacher14}. The procedure for data reduction is similar to what is described in detail in \cite{raimundo19a}. The only difference is the sky correction. As there are no sky exposures we used a dark region of the field of view and a sky mask to do the sky correction during the recipe \textsc{muse$\_$scipost}. The final cube has a spatial and spectral dimension of 321 $\times$ 312 $\times$ 3681 pixels, covering a field of view of 1 $\times$ 1 arcmin$^{2}$, which at the redshift of NGC 5077 corresponds to $\sim 13 \times 13$ kpc. The pixel spatial sampling is 0.2 $\times$ 0.2 arcsec$^{2}$ and the spectral sampling is 1.25 \AA. The wavelength range covered by the observations is 4750 - 9350 \AA\ and the spatial resolution of the data is estimated to be 0.8 arcsec from a star that is observed south west of the galaxy. The median spectral resolution across the field of view is $\sigma = 60 \pm 6$ km\,s$^{-1}$ (FWHM = 2.6 \AA) at 5577 \AA, $\sigma = 44 \pm 8$ km\,s$^{-1}$ (FWHM = 2.37 \AA) at 6923 \AA, and $\sigma =  37 \pm 4$ km s$^{-1}$ (FWHM = 2.42 \AA) at 8399 \AA\ based on the fitting of unblended sky lines \citep{raimundo19a}.

To increase the signal-to-noise (S/N) ratio of the data, we produce two spatially binned data cubes using the Voronoi binning technique described by \cite{cappellari&copin03}. The first data cube is binned based on the S/N level of the continuum at 5500 \AA, to a minimum S/N $\sim$ 50 per spatial pixel (i.e. S/N $= 40/\AA$) to determine the stellar kinematics. To determine the stellar kinematics we use the latest Python implementation of the Penalized Pixel-Fitting (pPXF) method (\citealt{vandermarel&franx93}, \citealt{cappellari&emsellem04}, \citealt{cappellari17}), with the full suite of stellar templates from MILES (\citealt{sanchez-blazquez06}, \citealt{falcon-barroso11}) to determine the stellar line of sight velocity distribution (for more details see \citealt{raimundo19a}). Since the stellar and gas spatial distributions are roughly orthogonal, we found that binning the data-cube based on the continuum S/N resulted in loss of spatial resolution in the regions of low stellar flux but still strong gas emission. Therefore we created a second spatially binned cube based on the S/N in the H$\alpha$ line measured in the pure gas emission cube (calculated by subtracting the best fit pPXF model for the stellar emission from the observed spectra). We use an approach similar to \cite{rosales-ortega12} but for the H$\alpha$+[N II] line complex. We identify a spectral band that covers the H$\alpha$+[N II] emission, with a width of 80 \AA\ centred at the peak of the H$\alpha$ emission lines, and two adjacent spectral bands that cover the continuum (i.e. line-free emission). We then define the signal in the emission features as the mean of the difference between the pixel flux values within the emission spectral band and the mean of the fluxes in the two continuum bands, divided by the square root of the number of spectral pixels in the emission band. The noise is determined from the MUSE pipeline data reduction which provides a cube with the variance in each spatial and spectral pixel. To determine the ionised gas dynamics we use the available feature in the latest implementation pPXF, that allow us to simultaneously fit the stellar features and emission lines using a set of gaussian templates. More details will be presented in Section~\ref{sec:gas_dynamics}.

\section{Results and Discussion}
\subsection{Stellar kinematics}
\label{sec:stellar_kinematics}
We show the results of the pPXF stellar kinematics analysis in Fig.~\ref{stellar_maps}. The stellar light distribution in the galaxy is relatively smooth, as can be seen from the light contours (integrated in the full MUSE spectral range) in the left panel of Fig.~\ref{stellar_maps}. The central and right panels of Fig.~\ref{stellar_maps} show the stellar line of sight velocity and stellar velocity dispersion, respectively. The stellar rotation velocity is low, with values up to $\sim$50 km/s. The stellar velocity dispersion is higher in the centre of the galaxy (250 - 300 km/s in the central 10 arcsec) and along the major axis, and lower along the minor axis of the galaxy. The region of high velocity dispersion coincides with the presence of a stellar kinematically distinct core (KDC), clearly seen in the flip of the direction of the velocity vector in the middle panel of Fig.~\ref{stellar_maps}. This KDC has also been seen by \cite{caon00} and we determine its diameter to be $12.6$ arcsec $\sim 2.8$ kpc. We calculate the kinematic position angle of the KDC and of the main stellar body of the galaxy using the method of \cite{krajnovic06} implemented in the routine \textsc{fit\_kinematic\_pa} as used by \cite{cappellari07}. The KDC is consistent with a misalignment of 180 degrees, i.e. counter-rotation, with a kinematic position angle (PA) of $209.5 \pm 22.5$ degrees compared with the kinematic PA of $17.5 \pm 2.5$ degrees for the main body of the galaxy. The high velocity dispersion observed in the core indicates the presence of random stellar motions along the line-of-sight, and is likely due to the large mass of NGC 5077, but it is also possible that the inferred high velocity dispersion is partially the result of the superposition of orbits from the co-rotating and counter-rotating stars.

\subsection{Stellar and gas misalignment}
\begin{figure}
\centering
\includegraphics[width=9cm]{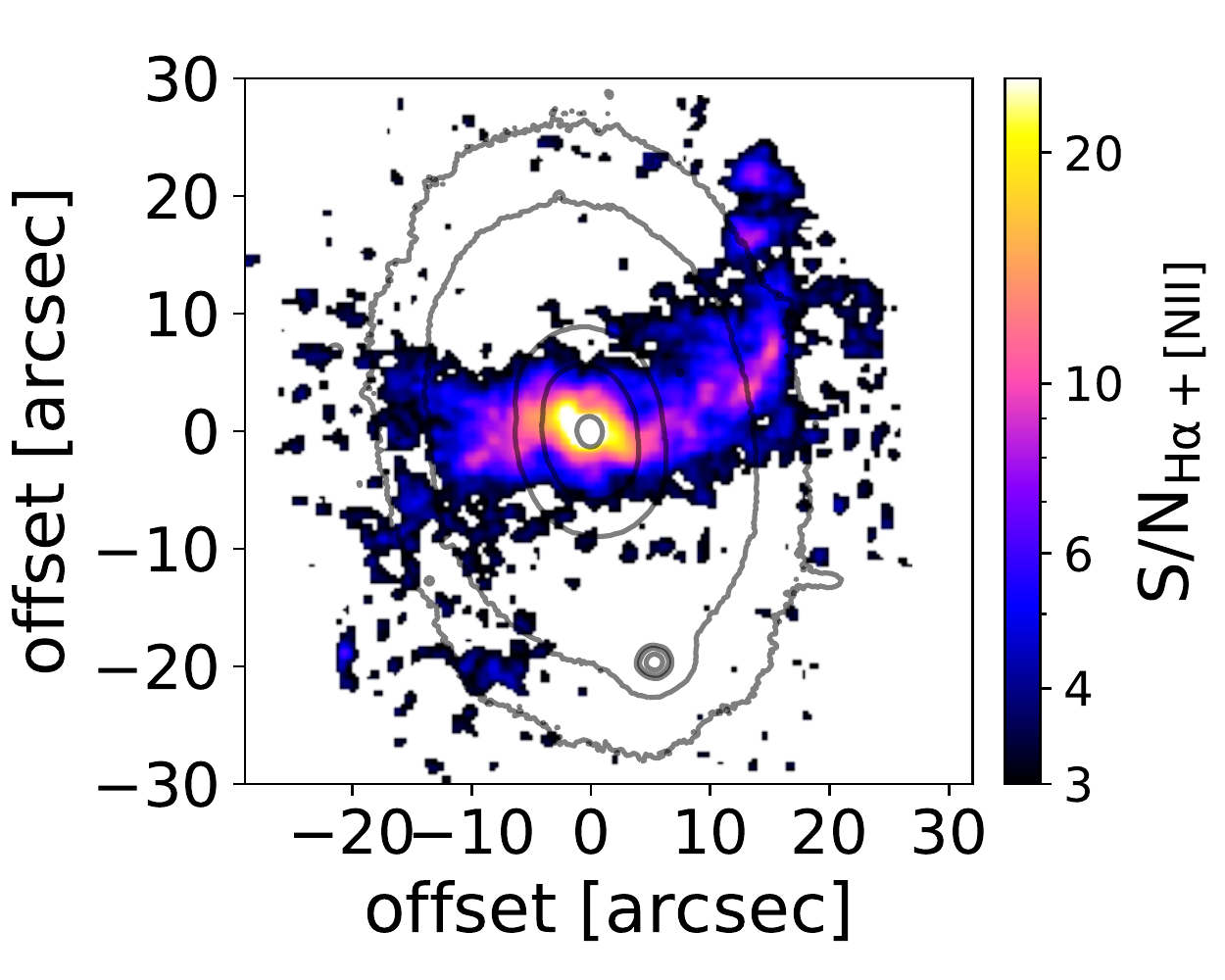}\\
\caption{Map of the signal-to-noise ratio per spatial pixel of the H$\alpha$ + [N II] emission in logarithmic scale. The S/N refers to the integrated continuum-subtracted spectrum in the waveband of H$\alpha$ + [N II] emission. The map illustrates the filamentary structure of the ionised gas distribution with the contours showing the isophotes of integrated light from Fig~\ref{stellar_maps}. North is up, east is left.}
\label{HalphaNII}
\end{figure}

Extended ionised gas emission is detected in NGC 5077 with a distribution that is significantly misaligned with respect to the main stellar body of the galaxy, and distributed along the galaxy's apparent minor axis, as also noted in previous work \citep{bertola91}. The MUSE data allows us to determine for the first time the extent of the gas distribution and the two-dimensional gas velocity map. In Fig.~\ref{HalphaNII} we show the ionised gas distribution as traced by the total emission in the wavelength region of the H$\alpha$ $\lambda$6563 and [N II] $\lambda$$\lambda$6548,6583 emission lines, similar to what was done to calculate the S/N of the emission features (Section~\ref{sec:data}). The map in Fig.~\ref{HalphaNII} shows the S/N of the total emission per unbinned spatial pixel within this wavelength band, with the spatial pixels with S/N $<3$ masked out. Fig.~\ref{HalphaNII} is mostly for visualisation purposes since it highlights pixels where significant signal is detected and illustrates the filamentary morphology of the ionised gas distribution. \cite{bertola91} estimate a total extent for the ionised gas as 30 $\times$ 15 arcsec. We find that the extent is actually almost twice as large, with ionised gas observed out to the edge of the field of view of our data cube, to a projected distance of $\sim$\,30 arcsec (6.5 kpc) from the nucleus. With the current data we cannot constrain if the emission extends further out beyond our field of view, as we are limited by the noise at the edge of the MUSE field-of-view and the size of the field-of-view itself. 

The strong misalignment between stellar and gas distribution can be clearly seen in Fig.~\ref{HalphaNII} where we overlap the contours for the integrated light on the ionised gas emission. The large scale gas distribution is filamentary and clumpy and distributed almost perpendicularly (in projection) to the stellar distribution. There are several changes of gas distribution PA as a function of radius: there is a change in PA at $r > \sim$4 arcsec from the centre when the gas becomes almost perpendicular to the major axis of the galaxy, and then at $r > 10$ arcsec. The gas emission for $r > 10$ arcsec starts following what appears to be tidal features seen in projection, with spiral-like regions of increased gas emission. The smaller radii regions have a lower PA offset from the stellar kinematic PA. In the west side the spiral arm is more pronounced, extending to the north and showing a filamentary structure surrounded by low level emission with an orientation parallel to the major stellar axis of the galaxy (as seen in projection). On the east side the emission is also filamentary and extends in two directions, a stronger component to the south and a weaker component with bright spots in the north-east direction. Previous work by \cite{caon00} detected extended gas emission but not the extended filamentary emission that we see in the N-S direction both east and west of the nucleus.

\begin{figure*}
\centering
\includegraphics[width=12cm]{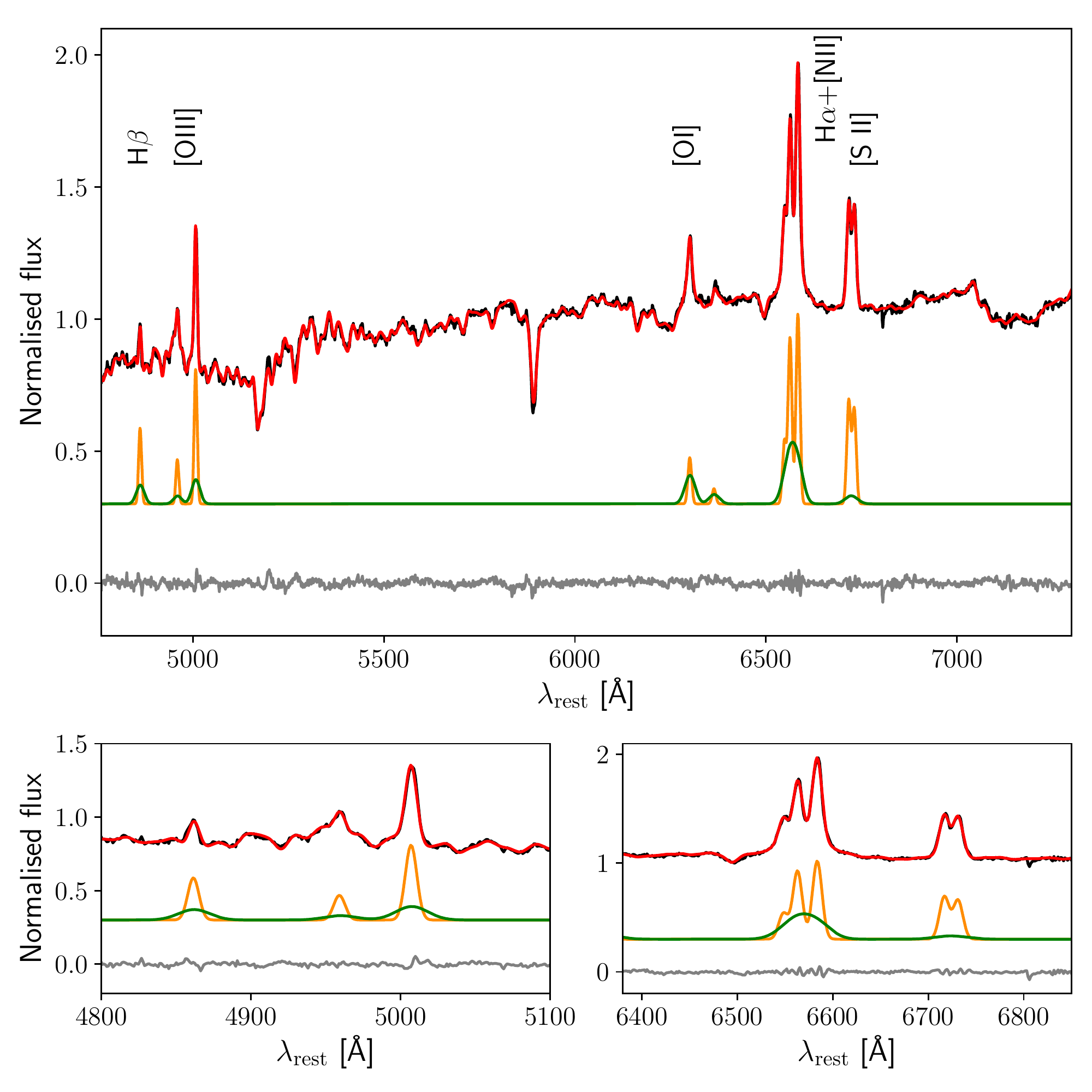}
\caption{Nuclear spectrum of NGC 5077 extracted from an integrated region of 1 $\times$ 1 arcsec$^{2}$ (black solid line). The region is centred at the position of the AGN as given by the peak of the H$\alpha$ integrated emission. The best fit model is shown in red. The two emission line components (green and orange) are shown in the middle of the panels, with a vertical offset for illustration purposes. The residuals from the fit are shown in grey in the bottom of each panel. The top panel shows a wide wavelength range while the two bottom panels show a zoom-in of the spectral region of the H$\beta$ and $[$O III$]$ emission lines (left) and the H$\alpha$, $[$N II$]$ and $[$S II$]$ emission lines (right). The emission lines are well represented by the presence of two kinematic components, one with velocity dispersion $\sim$ 210 km/s and another one with $\sim$ 640 km/s.}
\label{nuc_spec}
\end{figure*}

\subsection{Ionised gas dynamics}
\label{sec:gas_dynamics}
The galaxy spectra show several emission lines (H$\beta$ $\lambda$4861, [OIII] $\lambda$$\lambda$4959,5007, [OI] $\lambda$6300,6364, H$\alpha$ $\lambda$6563, [NII] $\lambda$$\lambda$6548,6583 and [SII] $\lambda$$\lambda$6716,6731) that trace ionised gas. We start by analysing the emission features in the nuclear spectrum. In Fig.~\ref{nuc_spec}, we show the nuclear spectrum (black solid line) and the result of the pPXF best fit model (red solid line) for that region. The individual emission lines are shown in orange and green and the residuals from the fit are shown in grey at the bottom of each panel. We find that in the nuclear region two sets of emission lines are needed to reproduce the observed spectrum. One of the components is relatively narrow (shown in orange in Fig.~\ref{nuc_spec}) and the second component is broader (shown in green), redshifted by $\sim$ 100 km/s with respect to the systemic velocity of the galaxy and with velocity dispersion of $\sigma = 640$ km/s (v$_{\rm FWHM} = 1500$ km/s). We find that this second component is present both in the Balmer and in the forbidden lines ([O III], [N II], [S II]), indicating that it is not a manifestation of the broad line region of the AGN but most likely of the complex kinematics and non-circular motions in the narrow line region. Previous work (\citealt{balmaverde&capetti14}, \citealt{cazzoli18}) have also found evidence of this second component. The detection of a broad line region component and hence the classification of NGC 5077 as a type 1 AGN is however, still being argued. While \cite{balmaverde&capetti14} find that there is no secure detection of a third component (associated with the H$\alpha$ emission from the broad line region) \cite{cazzoli18} find that the addition of a broad H$\alpha$ component reduces the residuals in their fit. In this work we find that two components are able to reproduce the emission lines in the nucleus of NGC 5077, as can be seen in Fig.~\ref{nuc_spec}. While a weak broad emission line component may be present, our spectra does not show strong evidence for emission from the broad line region. 

\begin{figure*}
\centering
\includegraphics[width=19.5cm]{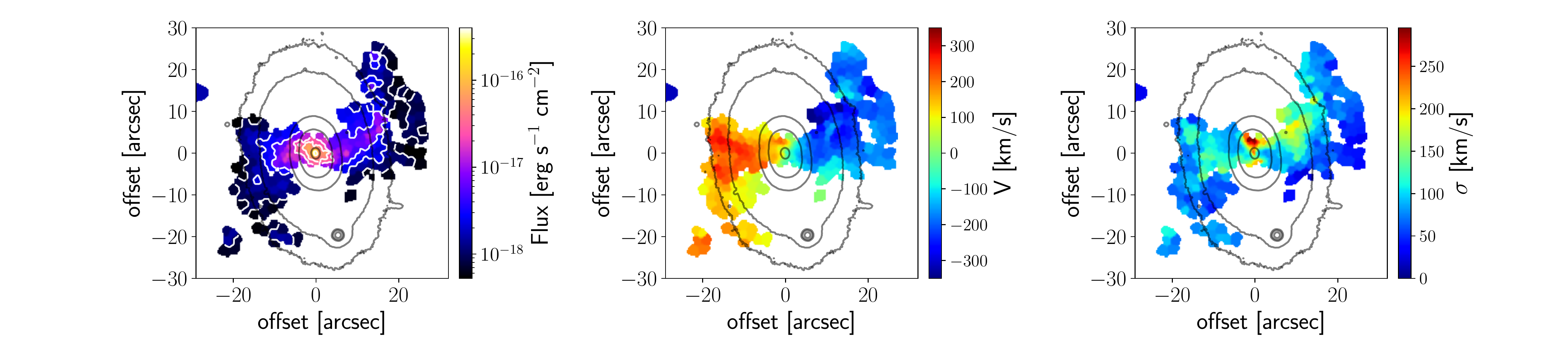}\\
\includegraphics[width=19.5cm]{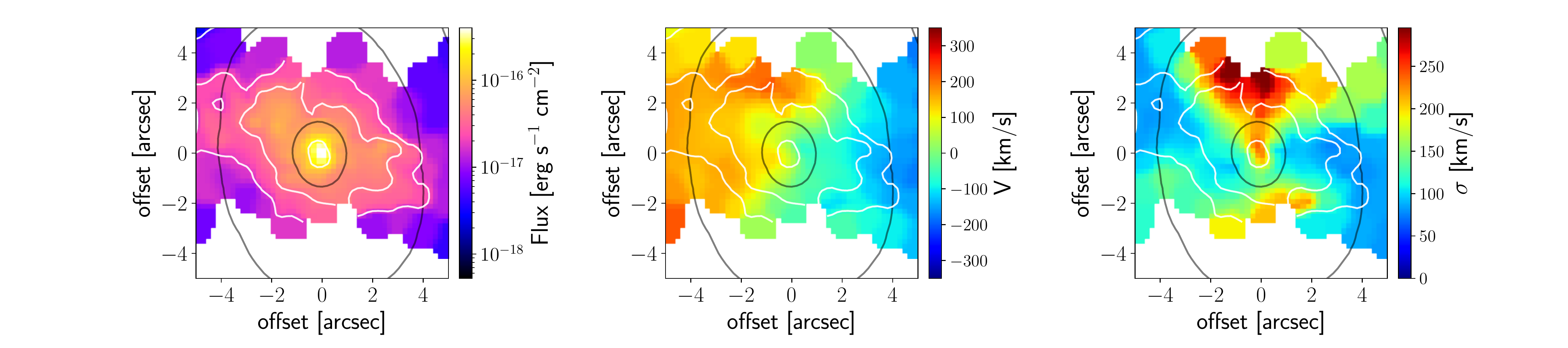}\\
\caption{Maps of ionised gas distribution and dynamics based on the H$\alpha$ narrow emission. Top panels show the full field-of-view and bottom panels a zoom-in of the inner 10 $\times$ 10 arcsec$^{2}$. 10 arcsec corresponds to 2.2 kpc. From left to right: H$\alpha$ flux, velocity and velocity dispersion. The flux value is per spatial pixel (0.2 $\times 0.2$ arcsec$^{2}$). The overlaid grey contours show the galaxy light isophotes while the white contours are H$\alpha$ flux contours. North is up, east is left.}
\label{Ha_map}
\end{figure*}

\begin{figure*}
\includegraphics[width=19.0cm]{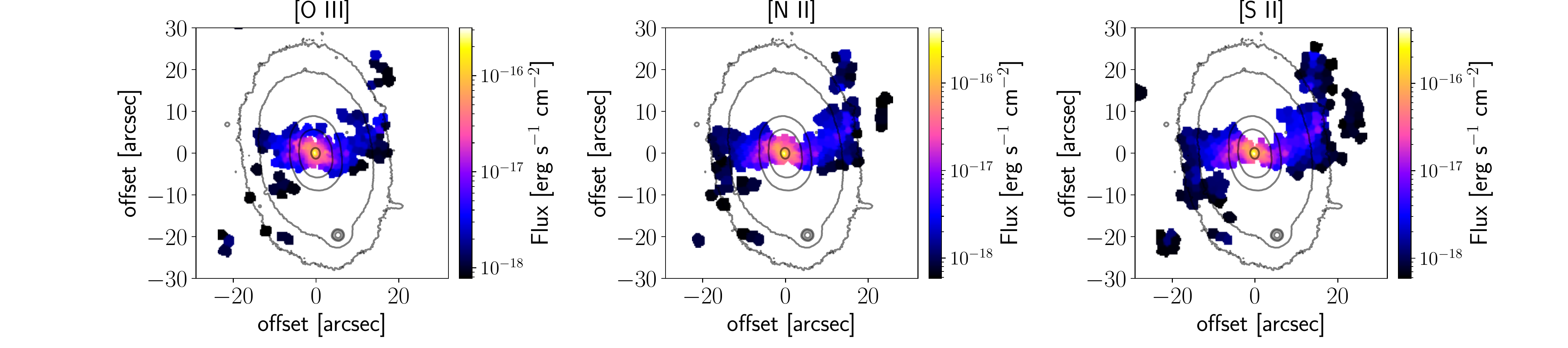}\\
\caption{Flux maps for [O III], [N II] and [S II]. The figures only show regions where each of the lines was detected at an amplitude to noise ratio A/N $ > 5$. The overlaid grey contours show the galaxy light isophotes. Orientation and scale is similar to that of Fig.~\ref{Ha_map}.}
\label{gas_maps}
\end{figure*}

To determine the overall gas dynamics in the galaxy we fit the full data-cube using pPXF. Since the stellar and gas distributions are not fully co-spatial, the size and spatial distribution of the bins for the two binned data-cubes described in Section~\ref{sec:data}, are different. We therefore adopt an approach similar to \cite{ganda06} to ensure that the stellar contribution is accurately determined in regions of strong gas emission but weak stellar emission. The bins in the data cube binned as a function of S/N in the continuum will be called `star bins' while the bins in the data cube binned as a function of S/N in the emission lines will be called `gas bins'. First we fit the stellar kinematics in the emission line binned data cube (i.e. data cube binned as a function of H$\alpha +$[N II] signal), using the best-fit stellar kinematics from Section~\ref{sec:stellar_kinematics} as starting values. The spectrum in each gas bin is calculated from the mean of the spectra in all the star bins that overlap with the gas bin. We use pPXF to determine the stellar kinematics in the gas bins, with the emission lines masked out of the fit. The stellar line of sight velocity moments from this fit are recorded and used as input for the subsequent analysis. 
The following step is to use the data cube binned as a function of H$\alpha +$[N II] signal and fit the stellar features and emission lines simultaneously. The stellar kinematics are held fixed to those determined in the step above, but the weights of each of the stellar templates and the additive Legendre polynomials are free to vary. We assume that the velocity and velocity dispersion of all the ionised gas species are tied and we use a single gaussian component per narrow line, with the flux of each line as a free parameter in the model. In the central regions of the galaxy we use a second, broader component for each line (as observed to be necessary in the spectrum of the nucleus - Fig~\ref{nuc_spec}). We also use a reddening curve (\citealt{calzetti00}, with R$_{\rm V} = 4.05$) to account for the presence of dust, with E(B-V)$_{\rm gas}$ as a free parameter in the fit and assuming that the intrinsic H$\alpha$/H$\beta$ flux ratio is 2.86 \cite{osterbrock89}.

\begin{figure*}
\centering
\includegraphics[width=12cm]{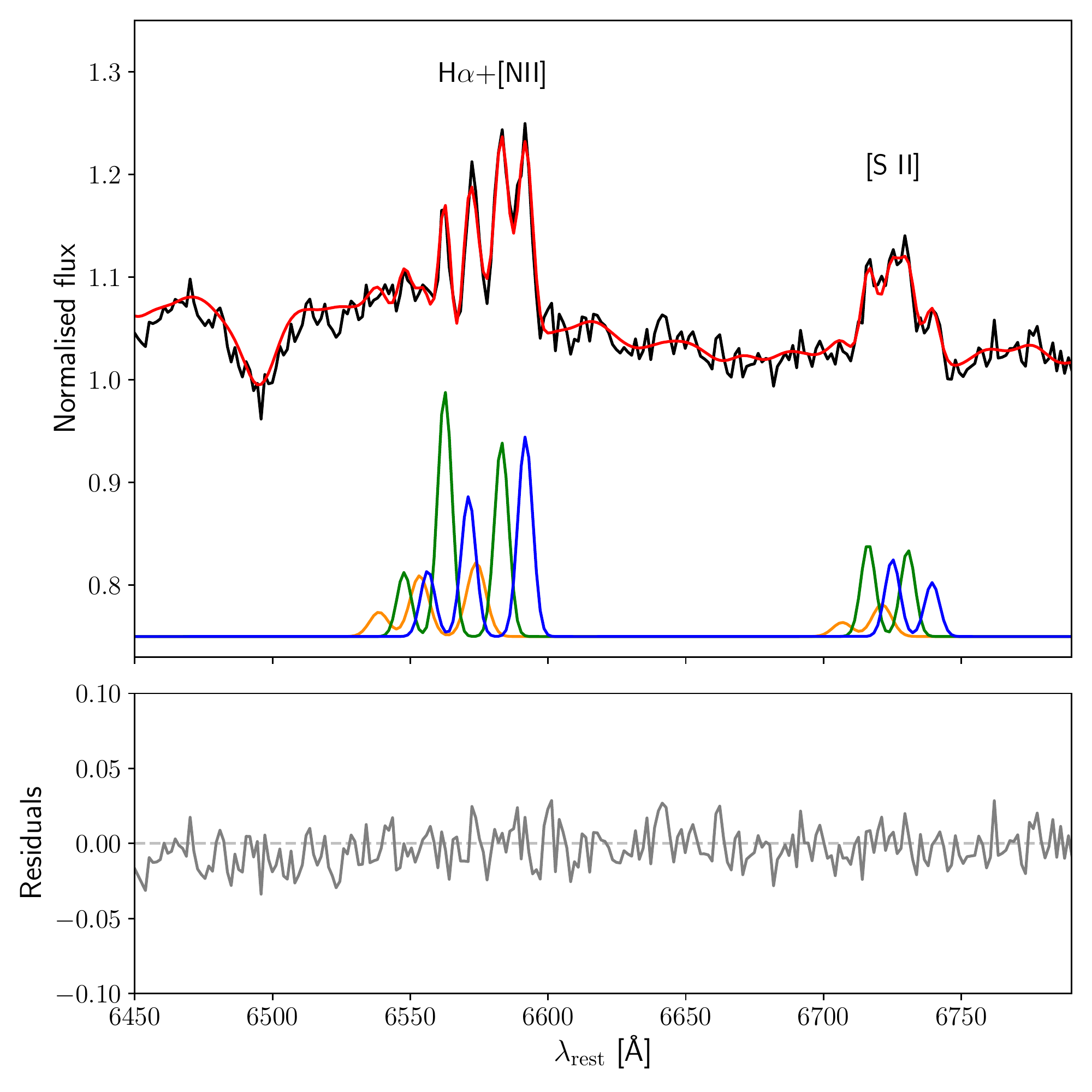}
\caption{Spectrum of the region of high velocity dispersion north of the nucleus, covering the H$\alpha$, [N II] and [S II] emission line complexes. To reproduce the emission, a model with three emission line components is necessary - see text for details.}
\label{north_spec}
\end{figure*}

In Fig.~\ref{Ha_map} we show the ionised gas velocity and velocity dispersion based on the results for the H$\alpha$ narrow emission line. The second broader component ($\sigma \sim$ 640 km/s) is only detected in the nucleus and we do not show it in the figures. A zoom-in of the central $10\times10$ arcsec$^{2}$ narrow line emission is shown in the bottom panels of Fig.~\ref{Ha_map}. The distribution of flux for the other species is shown in Fig.~\ref{gas_maps} and the line fluxes are corrected for reddening using the output from pPXF. Pixels where the amplitude to noise in each line is A/N $<$ 5, were masked out in the figures and all the maps shown were smoothed using a boxcar filter (running mean) of 2 $\times$ 2 pixels similar to the point-spread function (PSF) half-width at half maximum. 

The velocity map of the gas (Fig.~\ref{Ha_map}) shows redshifted velocities east of the nucleus and blue-shifted velocities west of the nucleus. The velocity profile is irregular in the central 8 arcsec along the major axis of the galaxy, a spatial region that coincides with the stellar counter-rotating core. As can be seen in the bottom panels of Fig.~\ref{Ha_map}, the line of zero velocity has an S-shape, with increased velocity dispersion (200 - 300 km/s) observed in the nucleus and in a cone-like shaped region towards the north out to a distance of r $\sim$ 3 arcsec, and towards the south, but with weaker emission. Increased velocity dispersion is also observed in two regions that extend out to 15 - 20 arcsec to the east and to the west, with $\sigma \sim 100 - 200 $ km/s. A close inspection of the spectra in these regions of higher velocity dispersion shows that the increased velocity dispersion is not caused only by a broadened emission line (due to turbulent motions for example) but by the superposition of more than one emission line component. It is likely that we are seeing different gas components along our line of sight, with slightly different velocities and locations. 

\begin{figure*}
\centering
\includegraphics[width=19.5cm]{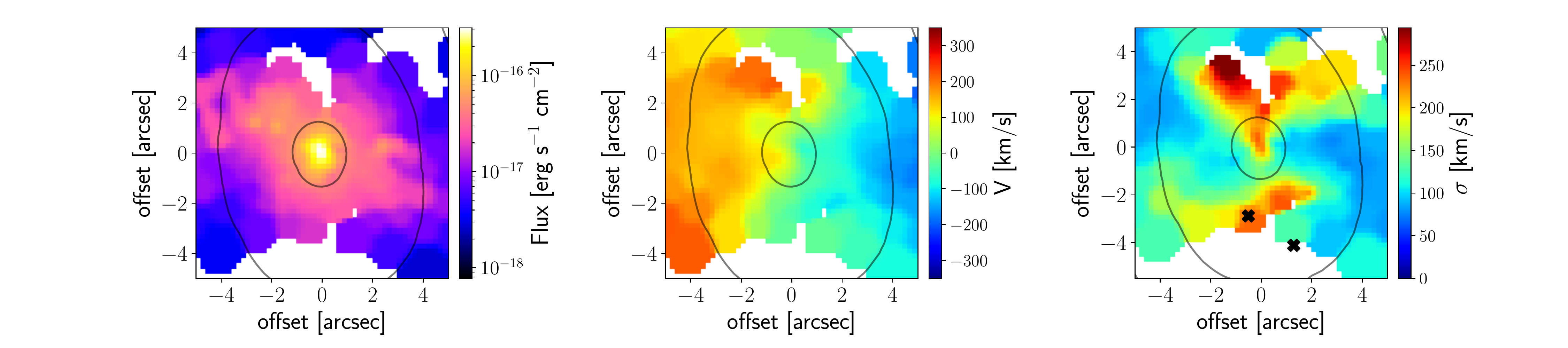}\\
\caption{Maps of the [O III] distribution and dynamics. Panels show a zoom-in of the inner 10 $\times$ 10 arcsec$^{2}$. 10 arcsec corresponds to 2.2 kpc. From left to right: [O III] flux, velocity and velocity dispersion. The overlaid grey contours show the galaxy light isophotes. The two black crosses in the right panel show the position of the two radio sources found by \cite{filho06}. North is up, east is left.}
\label{OIII_map}
\end{figure*}

Most of the gas is in the nucleus of the galaxy (r $< 10$ arcsec $\sim$ 2.2 kpc). However, we detect faint tails, separated by 180 degrees and approximately symmetric. In the north side we see hints of a double tail. Additionally we see regions of high gas velocity dispersion at r $> 5$ arcsec which are composed of double sets of emission lines, suggesting that we are seeing the superposition of two different gas components along our line of sight. Previous work suggested that the observed ionised gas lies in a warped polar ring \citep{bertola91}, which would give the gas emission a characteristic `S'-shape (e.g. \cite{albada82}). If that is the case, the tails would be caused by a warp in the gas ring/disc at larger radii, and the regions of high velocity dispersion would correspond to superposition of two different regions of the disc along our line of sight, caused by the warp.

\subsubsection{Nuclear gas outflow}
Fig.~\ref{north_spec} shows the spectrum of the region of high velocity dispersion north of the nucleus, corresponding to the integrated emission in a 1 $\times$ 1 arcsec$^{2}$ region, located 3 arcsec north of the nucleus. We model the spectrum from this region using a set of three gas components, which are necessary to reproduce the observed line emission. To verify that the data favours a model with three components instead of just two, we calculate the normalised residuals for the fit with two components and the fit with three components, given by (data - model) / 1$\sigma$ uncertainties, as in equation 13 of \cite{andrae10}.  We compare the distribution of residuals from each of the two models with a gaussian distribution with mean $\mu = 0$ and variance $\sigma^{2} = 1$. To evaluate which of the models is a better representation of the data (and therefore the one for which the residuals match a gaussian distribution best), we do a Kolmogorov-Smirnov test where we compare each of the model residuals to a gaussian distribution and calculate the respective p-values. We find that the p-value for the three-component model is p $= 7.346\times10^{-4}$ compared with p $= 2.8\times10^{-5}$ for the two-component model. The higher value of p-value for the three-component model means that this model is a better representation of the data, as its residuals more closely approximate a gaussian distribution than those of the two-component model.
One of the gas components (green line) has relatively low velocity (v $\sim 40$ km/s, $\sigma\sim$ 100 km/s), tracing a rotational component and matching the velocity of motion of the large scale gas distribution perpendicular to the major axis of the galaxy (Fig.~\ref{Ha_map}, bottom panel, centre). A second component (blue line) shows redshifted velocities of v $=$ 430 km/s and $\sigma\sim$ 110 km/s and a third component (orange line) shows blue-shifted velocities of v $=$ -380 km/s and $\sigma\sim$ 150 km/s. The blue-shifted component is mostly detected at the edges of the high velocity dispersion cone. In the south side of the high velocity dispersion cone, $\sim$ 2 arcsec south of the nucleus, the emission is dominated by the blue-shifted component (v $\sim$ -350 km/s) and the rotational component. The bicone geometry and the blue-shifted/redshifted emission, indicate that we are seeing an AGN driven outflow. The lack of a young stellar population and the geometry of the outflow disfavours a supernova driven outflow. Observing both blueshifts and redshifts at the same position along the line of sight and the more pronounced blue-shifted emission along the edges of the bicone are in agreement with models of AGN driven outflows along a hollow bicone that intercepts the plane of the sky (e.g. \citealt{bae&woo16}, \citealt{venturi17}). In Fig.~\ref{OIII_map} we show the flux, velocity and velocity dispersion of the one component fit to the [O III] emission in the nucleus where the outflow is more clearly visible.

\begin{figure*}
\centering
\includegraphics[width=15cm]{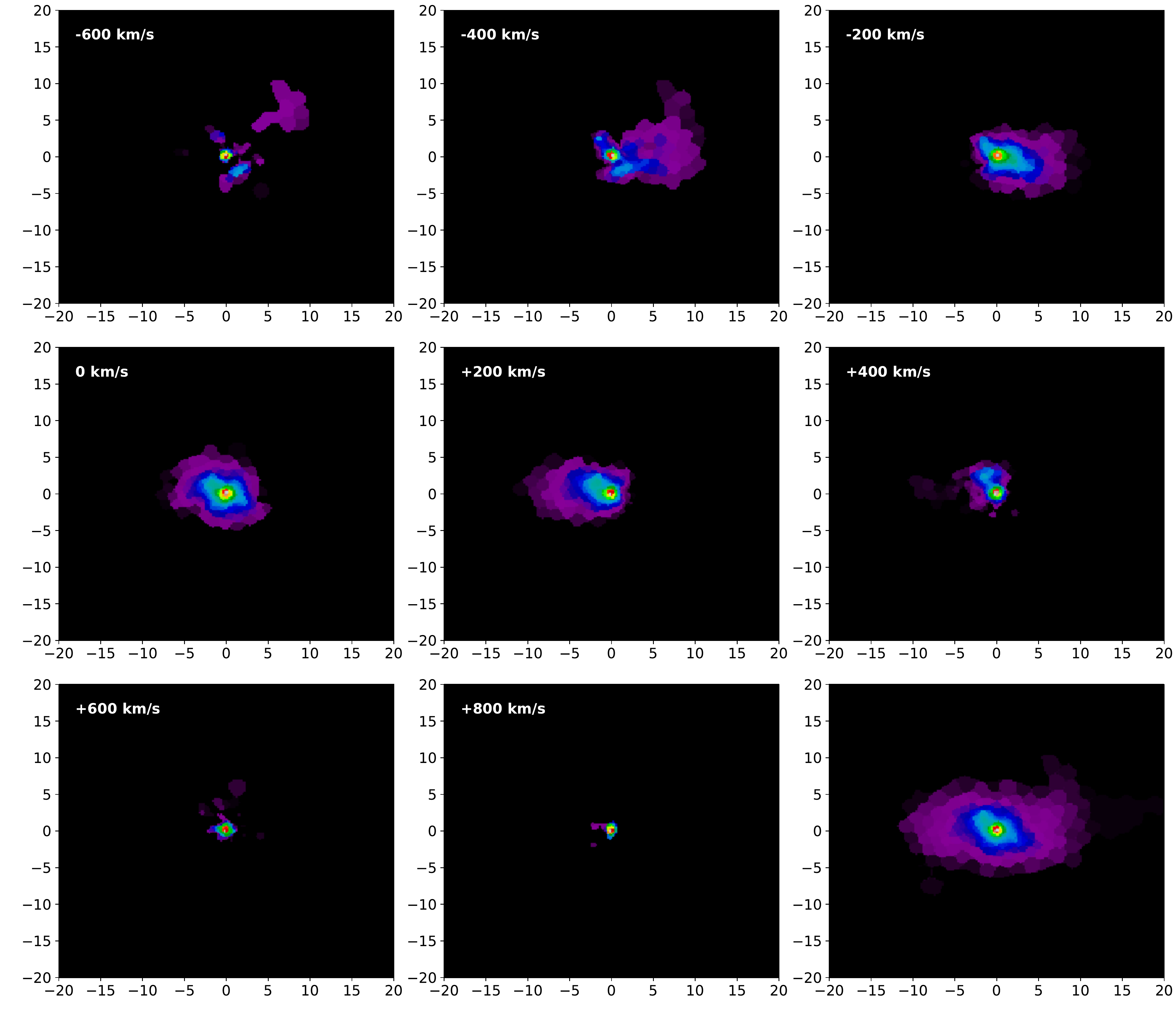}\\
\caption{Tomography analysis of the [O III] 5007 \AA\ emission. The maps show the [O III] emission integrated in velocity slices of 200 km/s each. Each panel shows the emission for a slice, with the central velocity of the slice (measured with respect to the systemic velocity), indicated in the top left corner. The last panel in the bottom right corner shows the summed emission from all the velocity slices. The images are in square root scale to highlight the high velocity emission from the nuclear regions. The size of each panel is 40 x 40 arcsec with the x and y axis labels showing the arcsec offset from the centre, as in Fig.~\ref{Ha_map}.}
\label{OIII_tomography}
\end{figure*}

Due to the complexity of the line emission and limited S/N, it is not possible to separate the line emission into individual emission components in every spatial bin. To show the velocity distribution of the gas we use the [O III] emission line, which has the advantage of not being affected by contamination from neighbouring lines, as opposed to [N II] and H$\alpha$. Fig~\ref{OIII_tomography} shows the [O III] integrated emission in velocity bins of 200 km/s, with the rest-frame 5007 \AA\ corresponding to velocity zero. Only emission above 2$\sigma$ is integrated to avoid adding noise. As can be seen in the panels of Fig~\ref{OIII_tomography}, the [O III] has a complex velocity structure. The broad emission in the nucleus can be clearly seen in all panels, extending from the -600 km/s to the 800 km/s velocity bin, in agreement with the large velocity dispersion and 100 km/s redshift detected in the nuclear gas emission (Section~\ref{sec:gas_dynamics}). The blue-shifted part of the outflow can be seen in the top left panel, in the regions of emission north and south of the nucleus and along the direction of the stellar counter-rotating core (PA $\sim$ 209 degrees). The redshifted part of the outflow is seen in the +400 km/s and +600 km/s panels, north of the nucleus. It is also clear that the extended emission to the west (seen in the -400 km/s and -200 km/s panels) has a higher absolute velocity than the extended emission to the east (seen in the +200 km/s panels), suggesting that the overall gas motion has a small blueshifted offset, which means that the gas velocity distribution is kinematically asymmetric (e.g. \citealt{chung12}) or that a gas component is located in the foreground of the galaxy, between the galaxy and the observer.

It is not clear if the broad velocity component ($\sigma \sim$ 650 km/s) detected in the nucleus is the base of the outflow. It is likely the case since it is redshifted by 100 km/s. If that is the case, the outflow accelerates from the nucleus until the position of maximum blue-shifted and red-shifted velocities observed in Fig~\ref{OIII_tomography}. Since the nuclear broad velocity component is mostly confined to a spatial region consistent with the size of the PSF, it is likely unresolved in our observations, and the high $\sigma$ could be due in part to complex dynamical motions in the central 100 parsecs of the galaxy.  

\begin{figure}
\centering
\includegraphics[width=10cm]{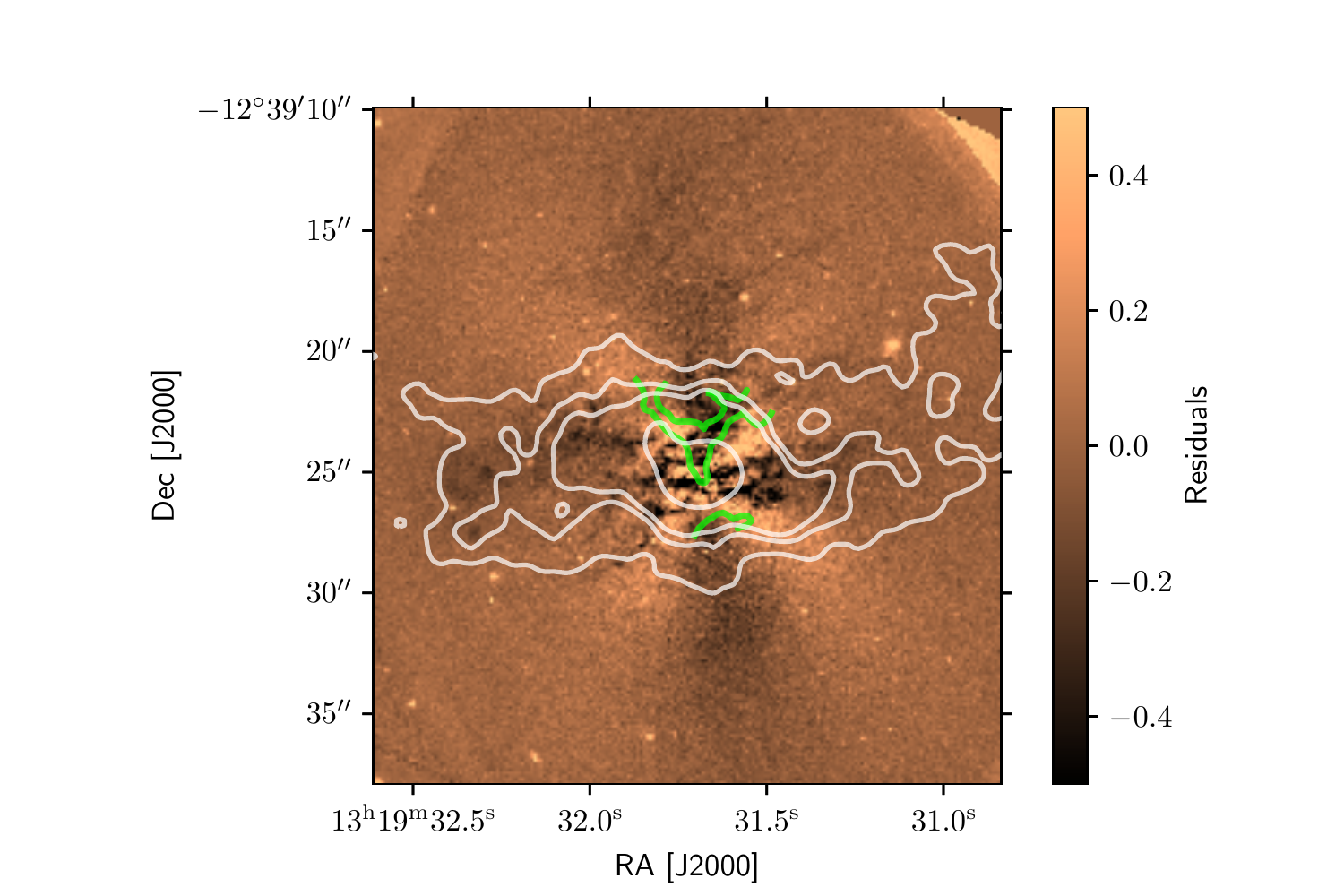}
\caption{Ellipse fitting residuals showing the filamentary dust on the HST WFPC2 F702W band image. The residuals with dark colours highlight regions of obscuration by dust. The white contours show the distribution of ionised gas while the green contours show the gas velocity dispersion levels of $\sigma = 200$ km/s and  $\sigma = 250$ km/s to show the direction of the AGN driven outflow. The dust follows the flux distribution of ionised gas in the east-west direction. In the north-south direction the dust is located in the same direction as the outflow.}
\label{ellipse_residuals}
\end{figure}

We compare the ionised gas emission with the dust distribution in NGC 5077. We use a \textit{Hubble Space Telescope} (HST)/WFPC2 archival image of NGC 5077 using the F702W filter to map the distribution of dust in the galaxy. Fig.~\ref{ellipse_residuals} shows the image residuals after subtracting isophotes from the HST image. The residuals with dark colours highlight regions of obscuration by dust and clearly show the filamentary nature of the dust in NGC 5077, consistent with a previous occurrence of an external accretion event. The white contours show the ionised gas distribution, and correspond to the levels of 5, 8, 10 and 25 of the S/N in the H$\alpha + $[N II] emission of Fig.~\ref{HalphaNII}. The green contours show the location of the highest velocity dispersion gas, corresponding to $\sigma =$ 200 km/s and $\sigma =$ 250 km/s in Fig.~\ref{Ha_map}. In the east-west direction, the dust distribution and the regions of higher ionised gas flux (shown with the white contours) are co-spatial, suggesting that they have a similar origin. There is an additional dust component along the major axis of the stellar light distribution (north-south) in the same direction as the outflow. In this direction, the inner regions of the outflow cone (shown by the green contours of increased velocity dispersion) are aligned with the north-south nuclear dust distribution. This can be more clearly seen in the region located at $\sim$ 2 - 3 arcsec north of the nucleus, where the northern part of the outflow overlaps with a region of dust absorption (shown by the dark brown colours in Fig.~\ref{ellipse_residuals}). Dust has been observed to be carried by outflows from the AGN (e.g. \citealt{baron18}) and it may be obscuring part of the outflowing gas emission, in particular in the south of the nucleus, where the outflow is not detected beyond 800 pc from the nucleus. 

The outflow cone must intersect the plane of sky, due to the observed blue-shifted and red-shifted emission corresponding to the front and back sides of a hollow outflow cone. The north part of the outflow is dominated by the redshifted emission while the south part of the outflow is dominated by the blue-shifted emission, which means that the south part of the outflow is pointing towards the observer. The material in the outflow likely originates in the disc of the galaxy and is being driven outwards by the AGN, meaning that the outflow cone partially intersects the disc of the galaxy. The gas and dust in the disc of the galaxy may be a remnant of the external interaction that created the counter-rotating stellar core.

There is a detection of two very strong radio sources in the nuclear regions of NGC 5077 using high spatial resolution observations \citep{filho06}. The authors note that without spectral information they cannot identify the origin of the radio emission (i.e. if it is coming from the AGN). We show the position of the radio sources overlaid with the [O III] velocity dispersion in the right panel of Fig.~\ref{OIII_map}. The coordinates for the radio emission are as given in Table 3 of \citealt{filho06}, while the [O III] reference frame is that of the MUSE datacube, which matches the coordinates of the HST images and Gaia DR3 source positions. \cite{filho06} find that these radio knots are extended, but considering the high spatial resolution of the radio observations ($\sim 0.1$ arcsec), we can see that the radio emission originates from outside the nucleus of the galaxy (i.e. the peak of gas emission and stellar light). The radio emission is spatially located in the region of the blue-shifted part of the outflow. In AGN driven outflows there is often a similar orientation of radio emission and outflow as we see here in NGC 5077, but not a perfect spatial correlation (\citealt{muller-sanchez11}, \citealt{fischer19}). The radio emission is in some cases associated with a jet that is interacting with the interstellar medium and transferring angular momentum to the gas to power the AGN outflow. In the case of NGC 5077, the radio emission may be caused by synchrotron emission from relativistic particles that are accelerated in the shocks caused by the AGN outflow (e.g. \citealt{zakamska&greene14}, \citealt{fischer19}). 

\subsection{Excitation mechanisms}
\label{sec:excitation}
\begin{figure*}
\centering
\includegraphics[width=18cm]{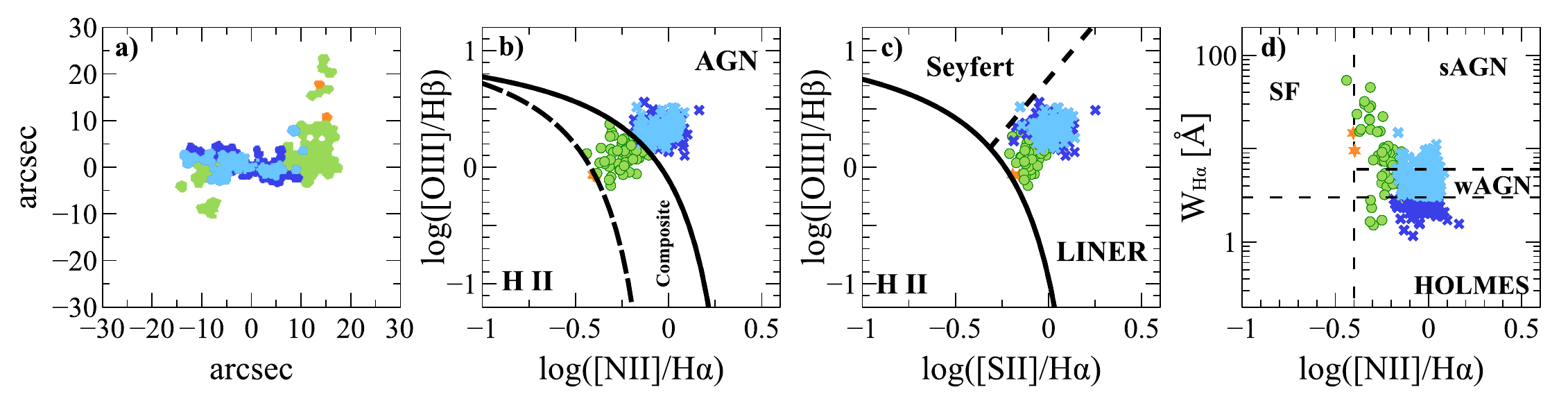}
\caption{BPT spatially-resolved diagrams for the regions of NGC 5077 where ionised gas (H$\beta$, [O III], H$\alpha$, [N II] and [S II]) is significantly detected. The left panel shows colour-coded regions to facilitate the identification of spatial trends. Panels b) and c) each show a different line ratio diagnostic and panel d) shows the H$\alpha$ equivalent width as a function of [N II]/H$\alpha$ (WHAN plot). The labels and solid/dashed lines indicate the plausible excitation mechanisms in each region of the diagram (see text for description). The coloured symbols match the line ratios measured to the corresponding colour-coded regions of the left panel. The diagrams show that there is a dominant contribution of AGN excitation in most of the regions (green circles and light blue crosses). The dark blue crosses show regions where post-AGB stars make an equal or larger contribution than AGN to the ionising field.}
\label{BPT}
\end{figure*}

We use the observed spatially resolved narrow emission line fluxes to investigate the excitation mechanisms at work in NGC 5077. The diagnostics we use are based on line flux ratios (\citealt{baldwin81}, \citealt{veilleux&osterbrock87}), also known as BPT diagrams (Fig.~\ref{BPT}). We only include spatial bins where all the diagnostic lines are detected above an amplitude to noise level A/N $> 3$. H$\beta$ is the weakest line and therefore the main factor in constraining the spatial regions where the BPT analysis can be carried out. In general, H$\beta$ and [O III] are much weaker than H$\alpha$ and [N II]. Panel a) of Fig.~\ref{BPT} shows the spatial regions analysed, colour coded to facilitate the reading of the BPT diagrams. Panels b) and c) show the [O III]$\lambda$5007/H$\beta$ vs [NII]$\lambda$6583/H$\alpha$ and the [O III]$\lambda$5007/H$\beta$ vs [SII]($\lambda$6716 + $\lambda$6731)/H$\alpha$ diagnostics, respectively. To identify excitation mechanisms based on these line ratios we use the theoretical regions defined by \cite{kewley06}, which are shown as solid and dashed lines in the panels b) and c) of Fig.~\ref{BPT}. 

As can be seen from the location of the points in panels a), b) and c) of Fig.~\ref{BPT}, the line ratios are consistent with the excitation conditions of LINERs for most of the regions in NGC 5077. Part of the outer regions (r $>$ 10 arcsec, 2.2 kpc) may also some contribution of star formation in addition to AGN, based on panel a). While typically LINER-like excitation is caused by a (low-luminosity) AGN, for some galaxies the LINER-like excitation is due to the presence of stars in the post-AGB phase (e.g. \citealt{stasinska08}, \citealt{singh13}, \citealt{levan17}). The argument is that in the post-AGB phase the stars become sufficiently hot to generate ionising photons that are able to produce LINER-like emission \citep{binette94}. If this is the case, it means that the stars responsible for the gas excitation must be older than 1 Gyr. 

As a second diagnostic we determine the equivalent width of the narrow component of the H$\alpha$ line (W$_{\rm H\alpha}$) to build a WHAN plot of W$_{\rm H\alpha}$ vs [N II]/H$\alpha$ (\cite{cidfernandes11}), which has shown to be a good discriminator between excitation by AGN and excitation by post-AGB stars \citep{cidfernandes10}. This diagnostic is also especially valuable for galaxies with weak lines \citep{cidfernandes11}, which is the case for H$\beta$ in NGC 5077. The equivalent width is determined from the H$\alpha$ flux and continuum of the spectral fitting, but similar results are obtained when integrating the narrow H$\alpha$ emission directly in the spectrum. Panel d) of Fig.~\ref{BPT} shows the equivalent width of H$\alpha$ (W$_{\rm H\alpha}$) as a function of the [N II]/H$\alpha$ ratio, with the dashed lines outlining the classification regions in Fig. 6 of \cite{cidfernandes11}. The left-hand side of panel d) shows the region where pure star forming galaxies (SF) are located, while sAGN and wAGN show where strong AGN (e.g. Seyferts) and weak AGN (e.g. LINERS) are located, respectively. The region labelled with `HOLMES' (Hot Low Mass Evolved Stars) indicates that there is a significant contribution of post-AGB stars to the gas excitation. This region is typically populated by galaxies that have stopped forming stars. Galaxies with AGN can still be located in the HOLMES region, but the ionisation photon output from the AGN is comparable or weaker to that of HOLMES \citep{cidfernandes11}.

From Fig.~\ref{BPT} we can infer that in NGC 5077 there is significant excitation by AGN throughout most of the field-of-view, including in the tidal arm to the north. The AGN activity is weak, characteristic of LINER, as can be seen from the location of all data points in panel c). Most of the LINER conditions are dominated by AGN excitation as opposed to post-AGB stars (panel d), except for some locations north and south of the nucleus and along the major stellar light distribution, where excitation by post-AGB stars makes a dominant contribution to the gas excitation (dark blue crosses.)
Additionally, there is further evidence of AGN (LINER-like) activity in NGC 5077 from the presence of nuclear X-ray emission with log L$_{\rm X (2-10 keV)}$ $= 6.8^{+1.6}_{-1.5} \times 10^{39}$ erg/s \citep{gultekin12}.

There are two other notable examples of S0 galaxies with decoupled extended ionised gas and an active nucleus, NGC 4993 and Mrk 3. NGC 4993 has a relatively weak AGN (L$_{\rm X\,(0.5 - 8 keV)}$ = $2\times 10^{39}$ erg s$^{-1}$), and \cite{levan17} argue that the line ratios observed are not caused by the AGN but by hot post-AGB stars or shocks, similar to what has been found in other early-type galaxies without AGN. Mrk 3, another S0 galaxy with ionised gas out to $\sim$ 5 kpc scales, also shows excitation by AGN \citep{gnilka20}, although the AGN in Mrk 3 is more luminous (L$_{\rm bol}$ = 2 $\times$ 10$^{45}$ erg\,s$^{-1}$) than NGC 5077. NGC 5077 appears to have a weak AGN but bright enough to dominate as the excitation mechanism. Alternatively, it is possible that the AGN in NGC 5077 was more luminous in the past.

\subsection{Stellar population}
\begin{figure*}
\centering
\includegraphics[width=19.5cm]{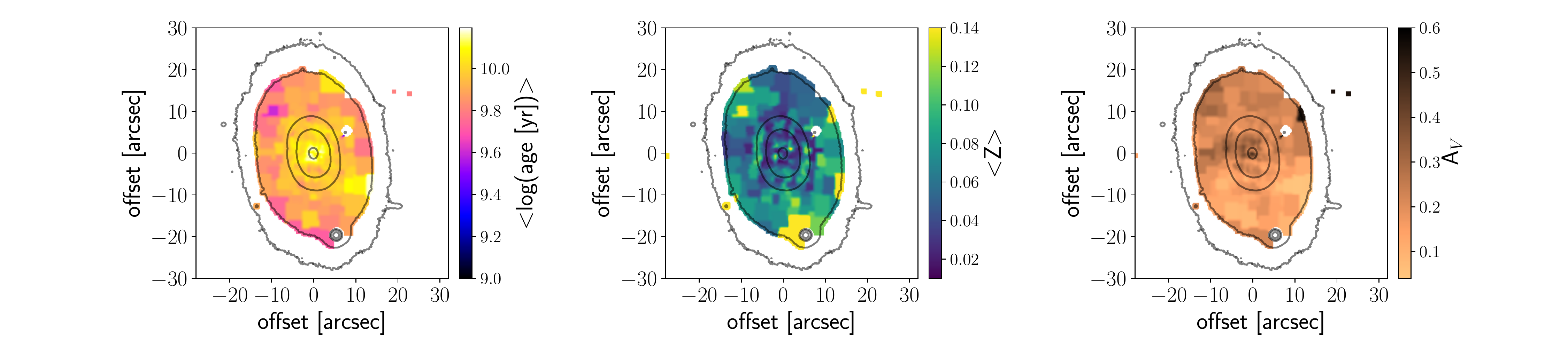}
\caption{Results from the stellar population synthesis modelling with \textsc{starlight}. Left: Luminosity weighted mean stellar age. Centre: Luminosity weighted mean metallicity. Right: V-band extinction (A$_{V}$). Integrated flux levels below 1 per cent of the peak flux were masked out of the image, as well as foreground stars.}
\label{starlight}
\end{figure*}

We use the stellar population synthesis code \textsc{STARLIGHT} (\citealt{cidfernandes05}, \citeyear{cidfernandes09}) to model the stellar population age and metallicity in the galaxy, to determine if there is evidence of recent star formation. We use the set of theoretical stellar spectral templates from the E-MILES evolutionary population synthesis models \citep{vazdekis16}, covering a range of stellar ages (0.0631 - 17.7828 Gyr) and metallicities (-2.32 $<$ [Fe/H] $<$ 0.22). The procedure for the modelling is similar to that described in \cite{raimundo19a} for the AGN host Mrk 590. More details can be found there and in \cite{cidfernandes04}, \cite{cidfernandes05} and references therein. We use as input the data cube binned to a minimum S/N of 50 and model the spectrum for each bin. We assume a Calzetti dust reddening curve \citep{calzetti00} to model the host galaxy reddening and \textsc{STARLIGHT} fits the V-band continuum extinction (A$_{V}$) at each spatial bin. The fitted wavelength range is 4644 $-$ 9000 \AA. Fig.~\ref{starlight} shows the results of the modelling: the luminosity-weighted mean stellar age ($<$log\,t$>$) in the left panel, the luminosity-weighted mean metallicity ($<Z>$) in the central panel and $A_{\rm V}$ on the right. Our results of an old stellar population in the nucleus of the galaxy agree with those found by \cite{annibali07}, who used long-slit spectroscopy to determine the stellar population within a radius of $2.85$ arcsec from the centre. We find that there is a slight negative gradient of stellar age outside the nucleus. We observe a younger stellar population with age $<$log$\,t>$ lower than 10 outside the nucleus. We note that the modelled stellar age and metallicity are less reliable in the outskirts of the galaxy and for this reason in Fig.~\ref{starlight} we only show the results from \textsc{STARLIGHT} for the central regions of the galaxy, where the stellar flux is greater than 1\% of the peak flux. There is no evidence of recent ($<$\,1 Gyr) star formation in the galaxy and the stellar population is dominated by old stars. From our analysis, the age of the stellar counter-rotating core is similar or slightly older than the stellar age in the main body of the galaxy, which points towards an early formation of the counter-rotating core. It is expected that the stellar population difference between kinematically distinct cores and the main body of the galaxy gets diluted by time, since the original accretion event \citep{mcdermid06}. This finding also supports the possibility that two external accretion events took place: one that formed the stellar counter-rotating core and a more recent one that resulted in the accretion of the decoupled distribution of ionised gas.

\subsection{Origin of the gas}
\label{sec:origin}
Due to the dynamical configuration observed, with the gas significantly misaligned and in a polar configuration with respect to the major axis of the galaxy, the origin of the gas can be clearly identified as external, as also seen for several other galaxies in similar configurations (e.g. \citealt{silchenko19}, \citealt{gnilka20}), as secular processes are not able to explain large quantities of misaligned gas (\citealt{caldwell86}, \citealt{kannapan&fabricant01}). 
The properties of NGC 5077 match rather well the predictions from numerical simulations by \cite{vandevoort15}, where an early type galaxy can show a misaligned gas distribution caused by late-time gas accretion after a gas-rich major merger. This is to the best of our knowledge the first time that the prediction of \cite{vandevoort15} of a warped misaligned gas disc is being confirmed by observations.

\cite{vandevoort15} carried out a numerical simulation as part of the FIRE simulations \cite{hopkins14}, of an early-type galaxy that underwent a major merger event (1:4 stellar mass ratio with a gas-rich galaxy). They predict that the merger will cause strong inflow of gas, followed by strong outflow where the gas is pushed out of the galaxy (and a significant fraction is actually expelled). Roughly 0.5 Gyr after the merger, the galaxy starts a late-time accretion of misaligned gas, leftover from the major merger and from minor satellites. Once the amount of accreted gas decreases, the gas starts its process of gravitational relaxation, and forms a misaligned warped disc (Fig. 1 of \citealt{vandevoort15}). The warp in the disc is caused by the gravitational torques of the stellar potential. In an axisymmetric potential, the gas will feel a gravitational torque from the stars, which is larger at smaller radii and therefore causes the realignment to happen faster in the nucleus. Since the gas is subject to dissipative forces, it will collide and reduce its angular momentum until it is in one of the most stable configurations, co-rotating or counter-rotating with the stars (e.g. \citealt{vandevoort15}). The result is that before the gas reaches a stable configuration, it will go through a relatively long phase ($\sim$ 2 Gyr) in a misaligned warped disc, with changes in the PA of the gas as a function of radius, due to the fact that the centre of the disc aligns faster than the outskirts. This is similar to what we observe in NGC 5077. The gas distribution in NGC 5077 is significantly misaligned from the stellar major axis, with the inner regions of the gas disc showing a smaller angular misalignment than the outer regions.
Below we summarise the observations of NGC 5077 that match what is predicted by the simulations of \citealt{vandevoort15}. 

The stellar counter-rotating core in NGC 5077 has an old stellar population and a large diameter, typical of the kinematically distinct cores found in slowly rotating galaxies \citep{mcdermid06}. Such KDCs are formed at early epochs (via a major merger or in-situ star formation after the accretion of gas) and no difference is expected to be seen in terms of the stellar population age \citep{mcdermid06}. Counter-rotating stellar cores, or KDCs can be long-lived and survive subsequent merger or gas accretion events \citep{ebrova20}. A KDC is also seen in the simulation of \citealt{vandevoort15}. It is then likely that the ionised gas distribution we see now has been formed from a more recent accretion event (after the major merger, as hypothesised by \citealt{vandevoort15}). This is also supported by the irregular gas velocity profiles we observe that indicate that at least part of the gas is not in equilibrium. 

The stellar distribution of the galaxy is smooth (Fig.~\ref{stellar_maps}), with a regular velocity field and currently shows no signs of having been disrupted by a gravitational interaction. We also do not detect a significant stellar component associated with the polar gas distribution, which would have been expected for the case of a recent major merger with another galaxy (e.g. \citealt{tal09}) or by strong tidal interactions. In the scenario of \citealt{vandevoort15}, this is expected since they estimate that the signatures of the merger (shells and tidal tails) disappear before the gas aligns with the disc. Gas misalignments are not expected to be associated with merger signatures in the stellar photometry. It is possible that there is a low luminosity diffuse stellar component that we do not detect, for example after one of the minor mergers, as stars in low-mass satellite galaxies can be easily torn during the merger and spread out into the stellar halo (e.g. \citealt{bullock&johnston05}).
Gas is more susceptible to perturbations due to its dissipative nature, and can be perturbed or transferred from another galaxy during a close encounter or accreted from nearby gas triggered by a flyby \citep{kannapan&fabricant01}, without necessarily involving stellar perturbations. 
The metallicity of the gas is also what we expect from a merger and not from infall of primordial gas through filaments. $[$N II$]$ is significantly detected in almost all regions where H$\alpha$ is also detected. The line ratio $[$N II$]$/H$\alpha$ is typically $> 0.5$ as can be seen in Fig.~\ref{BPT}, and oxygen lines are detected, which indicates that the gas is not primordial (e.g. \citealt{caldwell86}, \citealt{watkins18}). This suggests that what we are seeing is not infall of gas through a filament but accretion of gas from leftover material from a major merger event, transfer from a neighbour galaxy or the disruption and accretion of a low mass satellite galaxy. 

\subsubsection{Group environment}

\begin{figure}
\centering
\includegraphics[width=10cm]{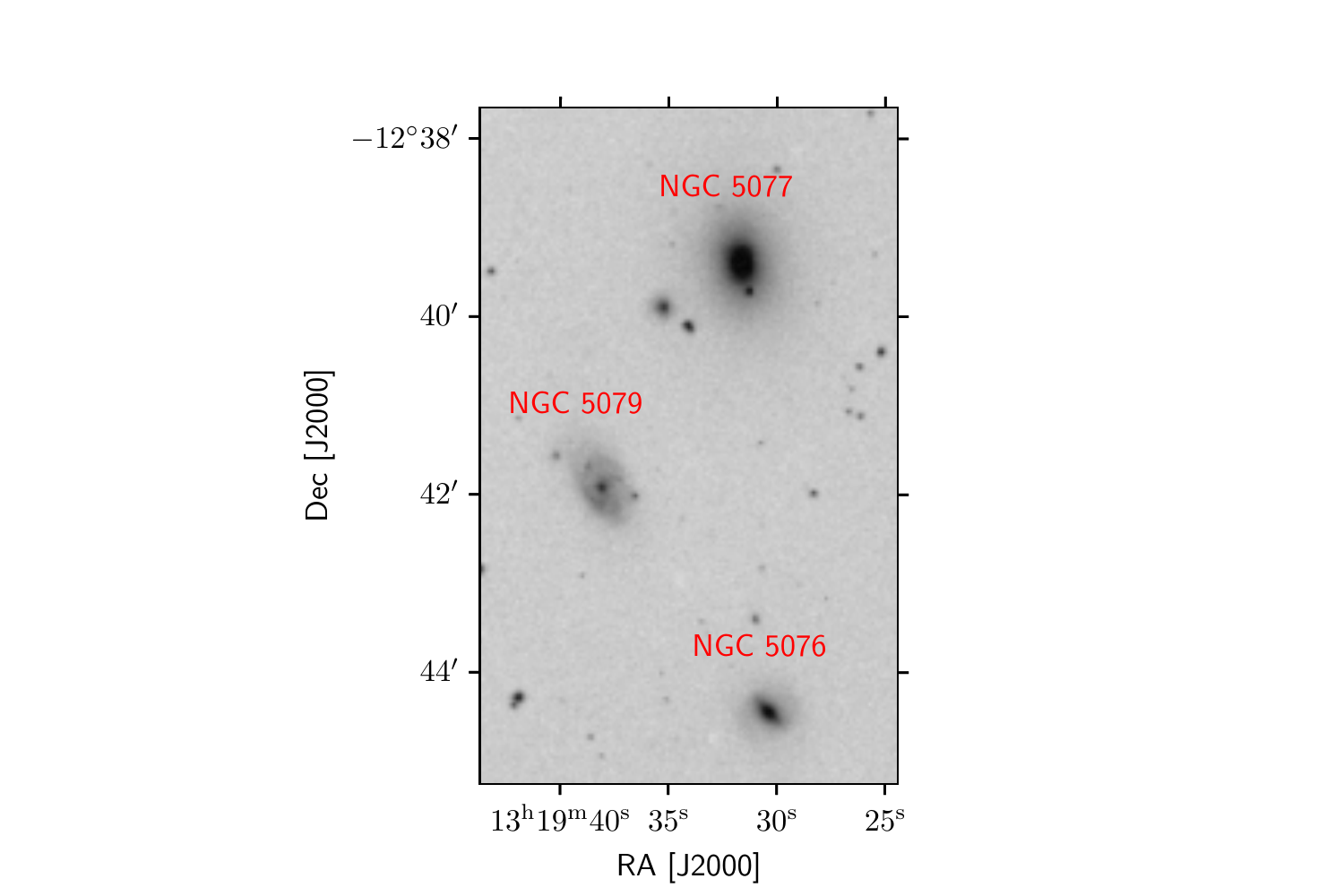}
\centering
\includegraphics[width=10cm]{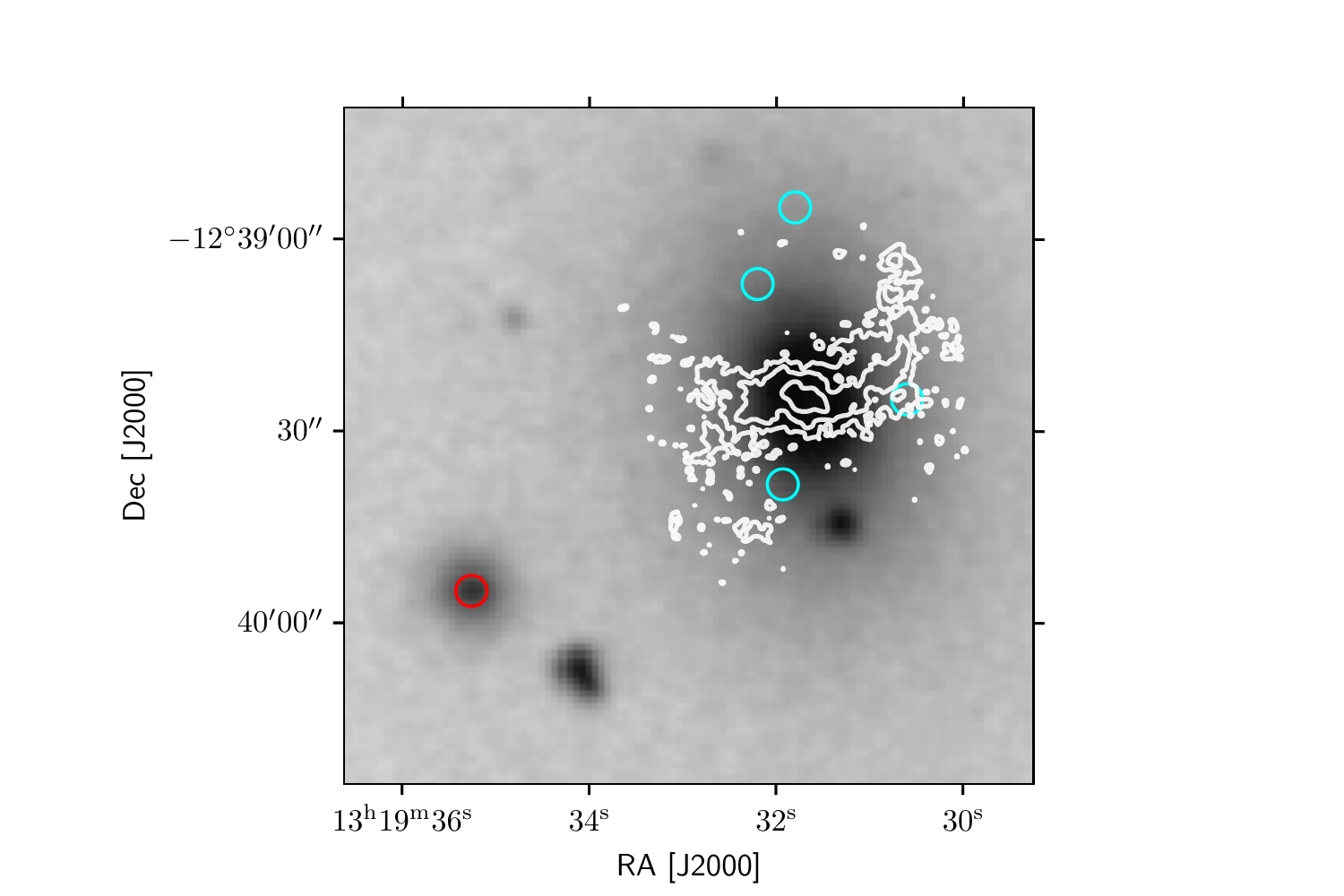}
\caption{I-band (6950 - 9000 $\AA$) images of the neighbouring region of NGC 5077 obtained from the Digitized Sky Survey (DSS2) catalogue. Top: Image of the Holm 514 galaxy group containing NGC 5077, NGC 5076 and NGC 5079. Bottom: Zoom in on NGC 5077. The red circle shows the position of Holm 514D one of the galaxies in Holm 514. The cyan circles show the positions of X-ray point sources detected by \textit{Chandra}. The white contours show the H$\alpha$ distribution in NGC 5077.}
\label{dss_neighbours}
\end{figure}

The simulated early type galaxy of \citealt{vandevoort15} is in a dense group, which acts as the source for the late time gas accretion (mainly through the accretion of satellites). NGC 5077 is also located in a group of galaxies which is not covered by the field-of-view of the MUSE data. In Fig.~\ref{dss_neighbours} we show the neighbouring region of NGC 5077, from a Digitized Sky Survey 2 (DSS2) I-band image. The top panel shows the group of galaxies Holm 514 of which NGC 5077 is a member, together with NGC 
5076 and NGC 5079. The bottom panel is a zoom in of the top panel, showing NGC 5077 with gas contours overlaid and a fourth group member, Holm 514D highlighted with a red circle. The other point sources are foreground stars. Based on the orientation of the tails of ionised gas, both Holm 514D and NGC 5079 are candidates for having provided some of the gas for NGC 5077 or for flybys that triggered the accretion of gas from the outskirts of NGC 5077 (leftover gas from a major merger for example). NGC 5079 is a barred spiral galaxy (SB(rs)bc pec \citealt{devaucouleurs91}) at a projected distance of 40 kpc and V$_{\rm sys} = 2242$ km/s while for NGC 5077 V$_{\rm sys} = 2806$ km/s. The estimated neutral hydrogen (H I) mass of NGC 5079 is $\sim 10^{9}$ M$_{\odot}$ \cite{bottinelli82} (although the source is possibly confused in their measurements). Radio observations at 1.4 GHz show that the emission in NGC 5079 is extended in the direction of NGC 5077 \citep{wrobel84}. In the case of spiral galaxies, the H I gas envelope is typically more extended than the stellar body, and that can explain why in tidal interactions or flybys the gas is stripped first (and not the stars) \citep{chung12}. Spatially resolved observations of neutral gas (H I) would allow us to constrain the origin of the gas, as gas transfer from companion galaxies typically shows H I bridges \citep{morganti06}. 

\begin{figure*}
\centering
\includegraphics[width=18cm]{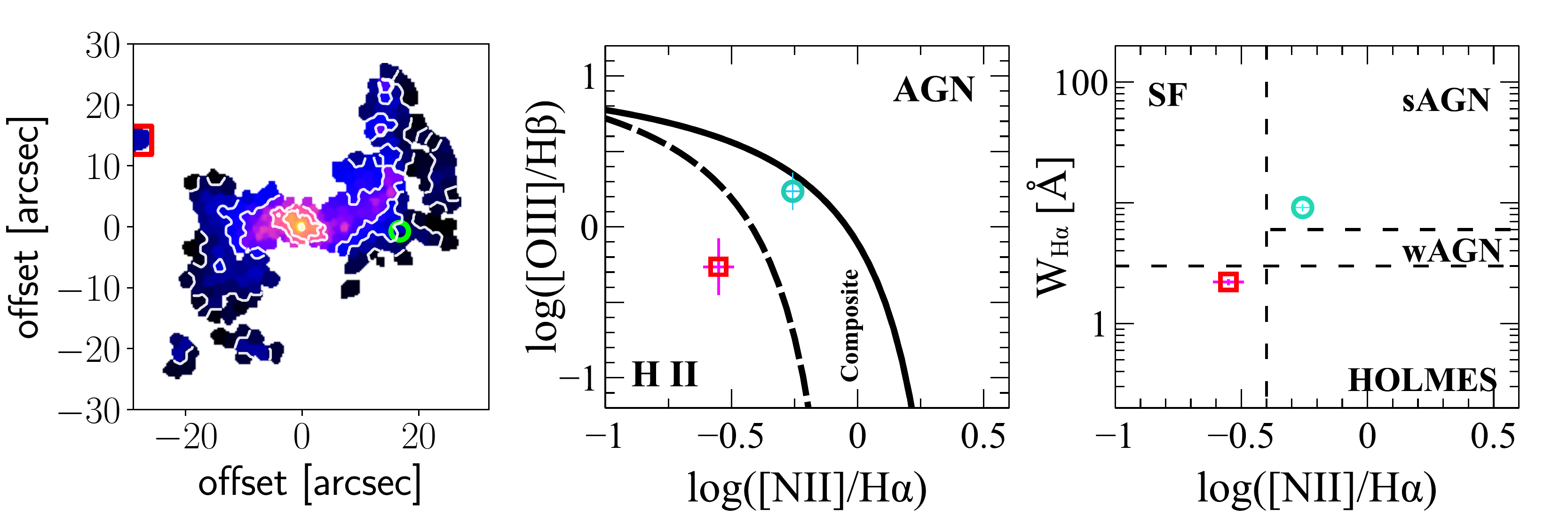}
\caption{Line diagnostics for two regions of NGC 5077: the H$\alpha$ blob to the north east and the X-ray point source west of the nucleus. The left panel shows the regions studied over-imposed on the H$\alpha$ flux map of Fig.~\ref{Ha_map}. The red square indicates the H$\alpha$ blob and the cyan circle the X-ray point source. The lines and labels in the central and right panels are similar to those in Fig.~\ref{BPT}. The H$\alpha$ emission region matches the line ratios expected for a star-forming region while the X-ray point source matches AGN excitation conditions.}
\label{BPT_cloud_xray}
\end{figure*}

The late time gas accretion could have originated from the accretion of one or more small gas-rich satellite galaxies, which have by now disappeared. That would explain the disturbance to the gas (but not the stars). There is an H$\alpha$ blob, a region of stronger H$\alpha$ emission at the edge of the MUSE field of view (shown as a red square in the left panel of Fig.~\ref{BPT_cloud_xray}). The excitation conditions are typical of star-forming regions, in contrast with the rest of the emission in NGC 5077 which is mostly consistent with LINER-like excitation conditions Fig.~\ref{BPT}). The region of strong H$\alpha$ emission appear to be at the end of the weak filaments in Fig.~\ref{HalphaNII}, that extend from the east side of the nucleus towards the north east edge of the field-of-view. This region is blue-shifted with respect to the systemic velocity of the galaxy, with $v = -310$ km/s, indicating that it is either in the foreground of the galaxy or moving towards the observer. It also shows very low velocity dispersion ($\sigma = 19$ km/s). It is not clear if the H$\alpha$ blob is a remnant of a disrupted gas-rich satellite that has been accreted by NGC 5077. As it is detected at the edge of the field-of-view, the low signal in the continuum does not allow us to detect a possible stellar population. Based on the H$\alpha$ flux in this region, we calculate a total H$\alpha$ luminosity L$_{H\alpha} = 4.5\times 10^{35}$ erg/s in a 220 pc $\times$ 220 pc region, which corresponds to a star formation rate of 2$\times 10^{-6}$ M$_{\odot}/$yr using the relation from \cite{kennicutt98}. Targeted observations of the outskirts of NGC 5077 and its companion galaxies (Holm 514D) would allow us to investigate the origin of the gas observed in NGC 5077.

The ionised gas distribution somewhat resembles NGC 4993, the host of the binary neutron star merger GW170817 an S0 galaxy that also underwent an external accretion event \citep{levan17}. Similar between these two galaxies is the decoupling between gas and stars, presence of dust lanes and the spiral-like features in the gas emission and the filamentary nature of the gas at large radii. We hypothesised that NGC 4993 may have had a similar history of external accretion as NGC 5077, possibly explained by the scenario of \cite{vandevoort15}. The S0 galaxy Mrk 3 \citep{gnilka20}, is another example of a galaxy with a large decoupled distribution of gas and the presence of dust lanes caused by accretion of gas from a companion galaxy. Such similarities point towards a common set of observables after an external accretion event in S0 galaxies.

\subsection{Black hole fuelling}
\label{sec:bh}

In terms of the consequences of this external accretion event to the black hole, we see that the external accretion event provided a fresh supply of gas to the galaxy. There is little gas co-spatial with the major stellar distribution, which indicates that even if there is some residual gas in the stellar disc, most of the gas in the galaxy has been externally accreted. Ionised gas is detected down to the spatial resolution of the data (FWHM $= 0.8$ arcsec, which corresponds to a radial distance of 0.4 arcsec $\sim 90$ pc from the nucleus) and shows misaligned motions. The nuclear gas appears to have a rotation component (Fig.~\ref{OIII_tomography}) in addition to the outflow. The dynamics of the gas shows that it is being affected by the stellar gravitational potential of the disc and losing angular momentum due to collisions, as it aligns itself in a co-rotating or counter-rotating orientation with respect to the stellar rotation. Accretion of misaligned gas can then be an efficient way to drive gas to the nucleus, and to promote the fuelling of the black hole. NGC 5077 has detections of prominent rotational H$_{2}$ lines in the mid infrared \citep{panuzzo11}, a tracer of molecular gas. The mass of gas available for black hole fuelling can be determined from future observations of the molecular gas mass in NGC 5077. 

In Fig.~\ref{dss_neighbours} we see that there are several X-ray point sources detected within the region of the galaxy (cyan circles), in addition to the X-ray emission coming from the nucleus of NGC 5077. While off-nucleus X-ray point sources may be caused by AGN in the background of the galaxy, the closer the X-ray point sources are to the centre of the galaxy, the more likely it is that they are associated with NGC 5077. There is one X-ray point source (CXOGSG J131930.6-123925), identified as part of NGC 5077 \citep{wang16}, that has ionised gas significantly detected at its location. The position of this off-nucleus X-ray point source is indicated by a cyan circle in the three panels of Fig.~\ref{BPT_cloud_xray}. From the contours in the left panel of Fig.~\ref{BPT_cloud_xray}, we can see that the position of the X-ray point source corresponds to a region of increased H$\alpha$ flux. The central and right panels of Fig.~\ref{BPT_cloud_xray} show the BPT and WHAN line diagnostic diagrams, which suggest that at the X-ray point source position (cyan circle), there are AGN-like excitation conditions. The following is speculation, but one hypothesis to explain both the presence of off-nucleus X-ray emission and the AGN-like excitation conditions detected at its position, is accretion of gas (possibly promoted by the recent inflow of fresh gas into the galaxy) onto an offset intermediate or supermassive black hole. Such offset black holes are expected in galaxies in which a minor or major merger occurred (e.g. \citealt{comerford09}), but it is not clear if that is what we are seeing in NGC 5077 and at the moment this is just a speculative hypothesis. Due to the low X-ray flux of the off-nucleus X-ray point source (L$_{\rm X}$ = 4.75$\times$10$^{38}$erg/s, \citealt{wang16}) it could also have a stellar origin (e.g. be a bright X-ray binary or supernova), and coincide with a region of increased H$\alpha$ flux and AGN-like excitation conditions.

Compared with MCG--6-30-15, the brightest AGN host with counter-rotating gas and stars \citep{raimundo17}, NGC 5077 appears to be at an earlier stage of the gas accretion process. In MCG--6-30-15 the ionised and molecular gas were aligned with the major axis of the galaxy but counter-rotating. The only large scale signature of interaction in MCG--6-30-15 was the presence of dust lanes. In NGC 5077 the accretion of gas is relatively recent and the gas appears to still be settling, as seen from the changes in PA as a function of radius. Curiously, MCG--6-30-15 also showed an AGN driven outflow. According to the scenario of \citealt{vandevoort15}, MCG--6-30-15 could be a later stage of the gas relaxation process that is currently occurring in NGC 5077. The AGN in MCG--6-30-15 is more luminous than NGC 5077, presumably having already been fuelled by the counter-rotating gas.

The large scale AGN-like excitation conditions in NGC 5077 can indicate that the black hole had a burst of activity in the past, possibly right after the major merger. The simulations of \citealt{vandevoort15} show that there was a strong inflow of gas due to the first major merger followed by an outflow that removed most of the gas from the galaxy. Some part of this gas could have fuelled the black hole for a short period of time. The misaligned gas that we currently see in NGC 5077 will continue to lose angular momentum and flow to the nucleus. The gas that we currently see in the nucleus may already be fuelling the black hole. It is expected that, as more gas reaches the nucleus, the black hole fuelling reservoir will significantly increase in mass (as seen in MCG--6-30-15) and fuel the future AGN activity in the galaxy.

\section{Conclusions}
Externally accreted gas can replenish the internal gas reservoir of early-type galaxies and may contribute to the fuelling of the black hole. To investigate the presence of counter-rotating structures in AGN host galaxies, we analysed observations of NGC 5077, an S0 active galaxy that shows an extended and dynamically decoupled distribution of ionised gas. Previous work has sampled the ionised gas along the galaxy's major and minor axes but this is the first time that the full extent and geometry of the ionised gas distribution is mapped and analysed, and that the origin of the misaligned gas is constrained.

We confirm the presence of a stellar kinematically decoupled core (KDC), close to counter-rotation with respect to the main body of the galaxy, and with a diameter of 2.8 kpc. The stellar age of the KDC is similar to that of the rest of the galaxy, indicating that it was not formed recently, as the difference in stellar age has been diluted. 
The galaxy shows low stellar velocity rotation ($v < 50$ km/s), and an increased stellar velocity dispersion in the KDC ($\sigma \sim $ 250 - 300 km/s). The gas distribution shows a strong warp and decoupled dynamics and is detected out to $\sim 6.5$ kpc. The gas velocity and velocity dispersion maps show complex dynamics and evidence for various gas components with slightly different velocities and positions along the line of sight. The gas PA changes as it approaches the nucleus, towards the direction of rotation of the stellar counter-rotating core. Gas is detected in the nucleus, down to the spatial resolution of our data at a radius of $\sim$90 pc from the central supermassive black hole. 
We detect an AGN driven outflow with maximum projected velocity of V $\sim 400$ km/s in the north-south projected direction. The outflow geometry is a hollow bicone that intersects the plane of the sky, with the south portion of the bicone pointing in the direction of the observer.
The dominant excitation mechanism of the ionised gas is AGN, based on the ionised gas line ratios and the H$\alpha$ equivalent width. The AGN in NGC 5077 is currently at a low luminosity level, log L$_{\rm X (2-10 keV)}$ $= 6.8^{+1.6}_{-1.5} \times 10^{39}$ erg/s \citep{gultekin12}, but the large physical extent of the ionised gas suggests that the AGN may have been more luminous in the past.

We present for the first time, observational evidence of the theoretical scenario proposed by \citealt{vandevoort15}, where the accretion of misaligned gas can create a warped gas distribution, with the central regions aligning first with the stellar rotation. Supported by the numerical simulations of \citealt{vandevoort15}, the most likely scenario for NGC 5077 is that several Gyr ago it went through a major merger with another galaxy, that resulted in the formation of the stellar counter-rotating core we currently observed. Most of the major merger gas was promptly removed temporarily or permanently from the galaxy by supernova driven outflows. Late time accretion (in the following 1 - 2 Gyr) of misaligned leftover gas from the merger (possibly triggered by flybys) or the accretion of small gas-rich satellites, provided fresh gas to the galaxy. When the accretion rate of gas decreased, the process of gas relaxation due to the stellar gravitational torques started to dominate, causing a significant warp to the gas, with the smaller radii gas aligning first. This is the stage that we currently observe in NGC 5077. It is expected that in the future, the gas will continue to lose angular momentum to end up in a stable configuration, i.e. co-rotating or counter-rotating in the same direction as the stars, similar to what is currently observed in MCG--6-30-15, an active galaxy with counter-rotating gas. The stellar torques to the gas can be an efficient method to drive gas to the nucleus and replenish the black hole fuelling reservoir.

NGC 5077 is part of an increasing number of active early-type galaxies in which the majority of the gas observed originated in an external accretion event. While a first look at the optical broadband image of NGC 5077 would not immediately identify the disturbances, ionised gas observations combined with the presence of a stellar distinct core clearly show that at least two accretion events have taken place. This highlights the importance of external accretion events for the gas supply in early type galaxies and to the replenishment of the reservoir of gas available for black hole fuelling. 

\section{Acknowledgements}
The author would like to thank the referee, for a thorough review of the paper and for the constructive comments and suggestions that have improved the quality of the paper. The author would also like to thank Mike Crenshaw, Ric Davies, Travis Fischer, Steven Kraemer, Joao Mendonca and Marianne Vestergaard for useful discussions. This project has received funding from the European Union$'$s Horizon 2020 research and innovation programme under the Marie Sklodowska-Curie grant agreement No 891744. Based on observations collected at the European Southern Observatory under ESO programme 094.B-0298. This research has made use of the services of the ESO Science Archive Facility. This research has made use of the NASA/IPAC Extragalactic Database (NED), which is operated by the Jet Propulsion Laboratory, California Institute of Technology, under contract with the National Aeronautics and Space Administration. Based on observations made with the NASA/ESA Hubble Space Telescope, and obtained from the Hubble Legacy Archive, which is a collaboration between the Space Telescope Science Institute (STScI/NASA), the Space Telescope European Coordinating Facility (ST-ECF/ESA) and the Canadian Astronomy Data Centre (CADC/NRC/CSA). This research made use of Astropy,\footnote{http://www.astropy.org} a community-developed core Python package for Astronomy \citep{astropy:2013, astropy:2018}. 

\bibliographystyle{aa} 
\bibliography{AGN} 

\end{document}